# Score and Information for Recursive Exponential Models with Incomplete Data.

Bo Thiesson*
Aalborg University / Microsoft Research
thiesson@microsoft.com

## Abstract

Recursive graphical models usually underlie the statistical modelling concerning probabilistic expert systems based on Bayesian networks. This paper defines a version of these models, denoted as recursive exponential models, which have evolved by the desire to impose sophisticated domain knowledge onto local fragments of a model. Besides the structural knowledge, as specified by a given model, the statistical modelling may also include expert opinion about the values of parameters in the model. It is shown how to translate imprecise expert knowledge into approximately conjugate prior distributions. Based on possibly incomplete data, the score and the observed information are derived for these models. This accounts for both the traditional score and observed information, derived as derivatives of the log-likelihood, and the posterior score and observed information, derived as derivatives of the log-posterior distribution. Throughout the paper the specialization into recursive graphical models is accounted for by a simple example.

## 1 Introduction

The recursive exponential models (REMs) of this paper have evolved from the recursive graphical models of Wermuth and Lauritzen (1983), which usually underlie the statistical modelling concerning probabilistic expert systems (Pearl 1988; Andreassen et al. 1989; Spiegelhalter et al. 1993) based on Bayesian networks.

For a recursive graphical model the structural relations between variables are represented by a directed acyclic graph, where each node represents a variable, $X_v$, and directed edges signify for each variable the existence of direct causal influence from variables represented by parent nodes, $X_{pa(v)}$. Markov properties with respect to the graph (Kiiveri et al. 1984; Lauritzen et al. 1990) imply that any distribution, which is structurally defined by the model, can be represented by tables of conditional distributions, $p(X_v \mid X_{pa(v)})$, which for each possible value of $X_{pa(v)}$ hold the local conditional distribution for $X_v$.

The REMs have evolved from these models by the desire to model the local conditional distributions in further detail. One may visualize the REMs as recursive graphical models, where local conditional distributions are defined by individual regular exponential models, denoted as *local models*. Any exponential model can be used as a local model, whereby it is possible to accomodate very sophisticated structural restrictions.

In situations it may happen that the same fragments of the model recur at different sites in the model, as is often the case in e.g. pedigree analysis. To deal with such cases, the REMs also allow different tables of conditional distributions to be modelled by a generic component.

A matching Bayesian interpretation of the modelling is considered. Besides the structural knowledge as specified by a given model, experts may also specify imprecise knowledge about parameters in the model. This knowledge can then be used for the construction of a conjugate prior distribution of parameters. The matching prior distribution for a REM factorizes into individual *local priors* associated for each local model. Hence, local priors can be considered independently.

In the general case the conjugate prior for a local exponential model is approximated by a multivariate normal distribution. If the local model is not restricted beyond being a probability distribution, the natural conjugate prior is defined by a Dirichlet distribution of probabilities.

In particular REMs can be used for the construction of Bayesian networks. A Bayesian network resembles a quantified model, that is, a particular distribution belonging to the set of distributions as defined by the model. Given a database of observations, the maximum likelihood estimate is the usual candidate for a such quantification. If expert knowledge on parameters is also available, the largest posterior mode becomes a natural alternative.

In situations of incomplete data, the determination of the maximum likelihood estimate or the largest posterior mode may call for iterative methods. The present paper has primarily been motivated by the need of providing the first and second order derivatives of the log-likelihood and log-posterior distribution to be used for iterative estimation methods, and for interfacing a sequential updating method (Spiegelhalter and Lauritzen 1990a, 1990b) to follow up on a quantified model as new observations occur. An application of first order derivatives for estimation with incomplete data in REMs is demostrated in a companion paper (Thiesson 1995). For recursive graphical models without local restrictions a similar derivation of first order derivatives was proposed in Spiegelhalter *et al.* (1993) and Lauritzen (1995) and given in Russell *et al.* (1995) with a gradient-descent application for estimation. In a study on methods for learning the structural relations between variables Chickering and Heckerman (1996) have applied the second order derivatives from this paper.

The first order and the negative second order derivatives of the log-likelihood are in the following denoted as the *score* and *observed information*, whereas the first order and the negative second order derivatives of the log-posterior distribution are denoted as the *posterior* score and observed information.

Section 2 defines the recursive exponential models. This includes the important notion of global and local variation independence of parameters, local exponential modelling, and how to relax the global variation independence to allow for parsimonious modelling of recurring fragments in the model. Section 3 and Section 4 derive the traditional score and observed information in the situation of a single incomplete observation. In Section 5 the expressions for the score and observed information are extended to apply for samples of independent observations. Section 6 covers the posterior score and observed information. It is suggested that a prior distribution obeys assumptions of relaxed global and local independence of parameters considered as random variables, which match the assumptions of variation independence. Section 7 investigates conjugate prior distributions and how to construct these from imprecise expert knowledge. Finally, Section 8 indicates further aspects of modelling. Annulment of local variation independence and so-called block recursive exponential models are proposed.

A simple extension of a recursive graphical model serves as a ongoing example throughout this paper.

## 2 Recursive exponential models

Let $X = X_V = (X_v)_{v \in V}$ be a finite set of classification variables, each defined on a finite set of levels $\mathcal{I}_v$. Let $A \subseteq V$, then $\mathcal{I}_A = \times_{v \in A} \mathcal{I}_v$ and the variables $X_A = (X_v)_{v \in A}$ take on values $x_A = (x_v)_{v \in A} \in \mathcal{I}_A$. For $A = V$ we omit the subscript. For a particular value $\theta \in \Theta$ of the parameter space $\Theta$ the joint distribution of $X$ is denoted $p(X \mid \theta)$, in which case the likelihood based on a complete observation $x \in \mathcal{X}$ is denoted $p(x \mid \theta)$. The number of parameters in a model, also called the dimension, is denoted $|\Theta|$. Likewise, $|\mathcal{I}_v|$ is the number of levels for $X_v$, and $|\mathcal{I}_A| = \prod_{v \in A} |\mathcal{I}_v|$ denotes the number of configurations of levels for $X_A$.

Given the recursive graphical structure, as argued in the introduction, a REM holds two assumptions on the parameter space, to be described below. Readers familiar with Spiegelhalter and Lauritzen (1990a, 1990b) and the line of work reported in Heckerman *et al.* (1995) may recognize the assumptions, as assumptions of variation independence between parameters in different local components of the model, which are used in these papers but not explicitly named.

By *global variation independence* the graphical structure reflects the assumption that any distribution of a given model factorizes into a product of conditional distributions, each parametrized by variation independent components of the total parametrization. That is,

$$p(X \mid \theta) = \prod_{v \in V} p(X_v \mid X_{pa(v)}, \theta_v), \qquad (1)$$

where $\Theta = \times_{v \in V} \Theta_v$, and $\theta_v \in \Theta_v$ completely specifies the relationship between the variable $X_v$ and its conditional set of variables $X_{pa(v)}$, and $p(X_v \mid X_{pa(v)}, \theta_v)$ denotes a table of conditional distributions, which holds a *local distribution* $p(X_v \mid \pi_v, \theta_v)$ for each parent configuration of levels $\pi_v \in \mathcal{I}_{pa(v)}$.

In some applications, particularly pedigree analysis, it is typical to restrain the tables of local conditional distribution by knowledge about some of these tables being equal. In order to allow this type of applications the global variation independence is relaxed in a certain way. Let $\tilde{v} \subseteq V$ specify a set of variables, which associate equal tables of conditional distributions, and denote by $\tilde{V}$ the total set of these equivalence classes. Equal tables must be parametrized by the same parameters. Hence,

$$p(X \mid \theta) = \prod_{\tilde{v} \in \tilde{V}} \prod_{v \in \tilde{v}} p(X_v \mid X_{pa(v)}, \theta_{\tilde{v}}), \qquad (2)$$

where $\theta_{\tilde{v}} \in \Theta_{\tilde{v}}$ specifies the relationship between $X_v$ and $X_{pa(v)}$ for any $v \in \tilde{v}$. If equal tables are represented by a single generic table, as will be assumed from now on, this is a more appropriate representation, which reflects the reduction of the parameter space into $\Theta = \times_{\tilde{v} \in \tilde{V}} \Theta_{\tilde{v}}$.

Let $(i_{\tilde{v}}, \pi_{\tilde{v}}) \in \mathcal{I}_{\tilde{v}} \times \mathcal{I}_{pa(\tilde{v})}$, where $\mathcal{I}_{\tilde{v}} \times \mathcal{I}_{pa(\tilde{v})} = \mathcal{I}_v \times \mathcal{I}_{pa(v)}$ for any $v \in \tilde{v}$, index a generic table. By *local variation independence* the parametrization of a table additionally factorizes into components associated for each local distribution in the table. Hence, by local variation independence $\Theta_{\tilde{v}} = \times_{\pi_{\tilde{v}} \in \mathcal{I}_{pa(\tilde{v})}} \Theta_{\tilde{v} \mid \pi_{\tilde{v}}}$, whereby the table of local distributions is assembled by probabilities $p(i_{\tilde{v}} \mid \pi_{\tilde{v}}, \theta_{\tilde{v} \mid \pi_{\tilde{v}}})$.

Now, consider the likelihood of a single observation $x$. The simplifying assumptions of variation independence between local components of parameters then has the effect of breaking this likelihood into the product of local likelihoods, where each local likelihood function is given by the distribution as picked from the appropriate generic table by setting $\pi_{\tilde{v}} = x_{pa(v)}$. Hereby, the likelihood of a single observation becomes

$$p(x \mid \theta) = \prod_{\tilde{v} \in \tilde{V}} \prod_{v \in \tilde{v}} p(x_v \mid x_{pa(v)}, \theta_{\tilde{v} \mid x_{pa(v)}}). \qquad (3)$$

This concludes the simplifying assumptions of variation independence. By lingering on the more pragmatic effect hereof concerning the issue of modelling, we notice that the assumptions break down the statistical modelling into more tractable local constructions of $\sum_{\tilde{v} \in \tilde{V}} |\mathcal{I}_{pa(\tilde{v})}|$ models for conditional distributions. These models are denoted as *local models*.

The statistical modelling by REMs does not stop at this point, though. To completely qualify as a REM, each local model must be structurally defined by a regular exponential model. The local likelihoods in (3) are therefore represented in the form

$p(x_v \mid x_{pa(v)}, \theta_{\tilde{v} \mid x_{pa(v)}})$
$= b(x_v, x_{pa(v)}) \exp\left(\theta'_{\tilde{v} \mid x_{pa(v)}} t_{\tilde{v} \mid x_{pa(v)}}(x_v) - \phi(\theta_{\tilde{v} \mid x_{pa(v)}})\right)$
$\qquad (4)$

where $'$ denotes the matrix transpose, $t$ denotes the statistics, $\phi$ the normalizing function, and $b$ the carrying density.

Readers familiar with the recursive graphical models may realize how the REMs relate to these models. Disregarding the possibility of specifying equal tables, the local exponential modelling makes the difference. A local model of a usual recursive graphical model is not restricted beyond the fact that it is a model of probability distributions. In contrast, the local exponential modelling by REMs allows sophisticated structural restrictions to be placed on each local model, if desired.

The following example illustrates the representation of an unrestricted local model in the framework of REMs.

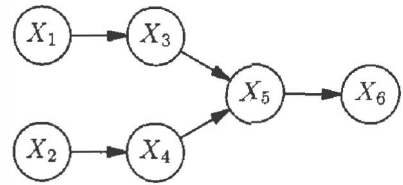

Figure 1: Graphical representation of the recursive response structure for the model considered in the example.

*Example:* Consider a model with recursive response structure as represented by the graph in Figure 1. Assume that this model obeys the assumptions of global and local variation independence except for domain knowledge, which dictates that the tables of conditional distributions are equal for the variables $X_3$ and $X_4$. In this case we loosen up the assumption of global variation independence by assuming that $\theta_{\tilde{3}} = \theta_3 = \theta_4$. Hence, $\theta = (\theta_1, \theta_2, \theta_{\tilde{3}}, \theta_5, \theta_6)$ and the likelihood of a single observation $x = (x_1, x_2, x_3, x_4, x_5, x_6)$ becomes

$p(x \mid \theta)$
$= p(x_1 \mid \theta_1) p(x_2 \mid \theta_2) p(x_3 \mid x_1, \theta_{\tilde{3} \mid x_1}) p(x_4 \mid x_2, \theta_{\tilde{3} \mid x_2})$
$\quad \times p(x_5 \mid x_3, x_4, \theta_{5 \mid x_3, x_4}) p(x_6 \mid x_5, \theta_{6 \mid x_5}).$

Consider variable $X_6$, which has a finite set of levels $\mathcal{I}_6 = \{i^0, i^1, \ldots, i^R\}$; the observed value $x_6$ being one of these. In this case, the last factor of the likelihood is picked from the local distribution $p(X_6 \mid x_5, \theta_{6 \mid x_5}) = (p^0, p^1, \ldots, p^R)$. Assume that the model for this distribution does not hold structural restrictions other than positivity constraints. An exponential representation of the local likelihood is then constructed as follows: Choose an index of reference, say $i^0$. Then, take as canonical parameters $\theta_{6 \mid x_5} = (\theta^1, \ldots, \theta^R)'$, where

$$\theta^r = \log \frac{p^r}{1 - \sum_{r=1}^{R} p^r} = \log \frac{p^r}{p^0},$$

and take as canonical statistics $t_{6 \mid x_5}(x_6) = (t^1(x_6), \ldots, t^R(x_6))'$, where

$$t^r(x_6) = \begin{cases} 1 & \text{for } x_6 = i^r \\ 0 & \text{otherwise.} \end{cases}$$

A minimal exponential representation is then given by

$$p(x_6 \mid x_5, \theta_{6|x_5}) = \exp\left(\theta'_{6|x_5} t_{6|x_5}(x_6) - \phi(\theta_{6|x_5})\right),$$

with normalizing function

$$\phi(\theta_{6|x_5}) = \log\left(1 + \sum_{r=1}^{R} \exp(\theta^r)\right).$$

□

## 3  Score of incomplete observation

Suppose for a given model that a complete observation $x$ is only observed indirectly through the *incomplete* observation $y$. The observation may be incomplete due to *missing* values according to a complete observation on $X$ or due to *imprecise* values. An imprecise value appears if the collector of data cannot distinguish between a set of possible values for a variable and therefore reports this set instead of a single value. Denote by $\mathcal{X}(y)$ the set of possible completions that are obtainable by augmenting the incomplete observation $y$. Under the condition that the observation is incomplete in an un-informative way the likelihood for the incomplete observation becomes

$$\begin{aligned} p(y \mid \theta) &= \sum_{x \in \mathcal{X}(y)} p(x \mid \theta) \\ &= \sum_{x \in \mathcal{X}(y)} \prod_{\tilde{v} \in \tilde{V}} \prod_{v \in \tilde{v}} p(x_v \mid x_{pa(v)}, \theta_{\tilde{v}|x_{pa(v)}}), \end{aligned} \quad (5)$$

where $\theta = (\theta_{\tilde{v}|\pi_{\tilde{v}}})_{\tilde{v} \in \tilde{V}, \pi_{\tilde{v}} \in \mathcal{I}_{pa(\tilde{v})}}$ denotes the vector of all parameters for the model.

Let $S(y \mid \theta) = \frac{\partial}{\partial \theta} \log p(y \mid \theta)$ denote the score of an incomplete observation. In accordance with the partitioning of the parameter vector into variation independent components, $\theta_{\tilde{v}|\pi_{\tilde{v}}}$, we describe the score by local components of dimension $|\Theta_{\tilde{v}|\pi_{\tilde{v}}}|$ given by

$$\begin{aligned} S_{\tilde{v}|\pi_{\tilde{v}}}(y \mid \theta) &= \frac{\partial}{\partial \theta_{\tilde{v}|\pi_{\tilde{v}}}} \log p(y \mid \theta) \\ &= \frac{1}{p(y \mid \theta)} \frac{\partial}{\partial \theta_{\tilde{v}|\pi_{\tilde{v}}}} p(y \mid \theta) \\ &= \frac{1}{p(y \mid \theta)} \sum_{x \in \mathcal{X}(y)} \frac{\partial}{\partial \theta_{\tilde{v}|\pi_{\tilde{v}}}} p(x \mid \theta), \end{aligned} \quad (6)$$

where the last equality follows from (5).

Consider the local derivatives of the likelihood for a complete observation. As

$$\begin{aligned} \frac{\partial}{\partial \theta_{\tilde{v}|\pi_{\tilde{v}}}} &p(x_v \mid x_{pa(v)}, \theta_{\tilde{v}|x_{pa(v)}}) \\ &= p(x_v \mid x_{pa(v)}, \theta_{\tilde{v}|x_{pa(v)}}) \\ &\quad \times \frac{\partial}{\partial \theta_{\tilde{v}|\pi_{\tilde{v}}}} \log p(x_v \mid x_{pa(v)}, \theta_{\tilde{v}|x_{pa(v)}}) \end{aligned}$$

the chain rule for differentiation implies

$$\begin{aligned} \frac{\partial}{\partial \theta_{\tilde{v}|\pi_{\tilde{v}}}} &p(x \mid \theta) \\ &= \sum_{v \in \tilde{v}} \left[ \chi^{\pi_{\tilde{v}}}(x_{pa(v)}) \frac{p(x \mid \theta)}{p(x_v \mid x_{pa(v)}, \theta_{\tilde{v}|x_{pa(v)}})} \right. \\ &\quad \left. \times \frac{\partial}{\partial \theta_{\tilde{v}|\pi_{\tilde{v}}}} p(x_v \mid x_{pa(v)}, \theta_{\tilde{v}|x_{pa(v)}}) \right] \\ &= p(x \mid \theta) \sum_{v \in \tilde{v}} \chi^{\pi_{\tilde{v}}}(x_{pa(v)}) \frac{\partial}{\partial \theta_{\tilde{v}|\pi_{\tilde{v}}}} \log p(x_v \mid \pi_{\tilde{v}}, \theta_{\tilde{v}|\pi_{\tilde{v}}}), \end{aligned} \quad (7)$$

where $\chi^{\pi_{\tilde{v}}}(x_{pa(v)})$ is the indicator function

$$\chi^{\pi_{\tilde{v}}}(x_{pa(v)}) = \begin{cases} 1 & \text{for } x_{pa(v)} = \pi_{\tilde{v}} \\ 0 & \text{otherwise.} \end{cases}$$

Thus, by the exponential representation (4) of the local likelihood for a complete observation, the partial derivatives of the likelihood become

$$\frac{\partial}{\partial \theta_{\tilde{v}|\pi_{\tilde{v}}}} p(x \mid \theta) = p(x \mid \theta) \sum_{v \in \tilde{v}} \chi^{\pi_{\tilde{v}}}(x_{pa(v)}) \left(t_{\tilde{v}|\pi_{\tilde{v}}}(x_v) - \tau(\theta_{\tilde{v}|\pi_{\tilde{v}}})\right), \quad (8)$$

where $\tau(\theta_{\tilde{v}|\pi_{\tilde{v}}}) = \mathbf{E}[t_{\tilde{v}|\pi_{\tilde{v}}}(X_v) \mid \pi_{\tilde{v}}, \theta_{\tilde{v}|\pi_{\tilde{v}}}]$, the expected value of the canonical statistic for the conditional distribution $p(\cdot \mid \pi_{\tilde{v}}, \theta_{\tilde{v}|\pi_{\tilde{v}}})$.

By inserting (8) into the expression for the local score of an incomplete observation, as given in (6)

$$S_{\tilde{v}|\pi_{\tilde{v}}}(y \mid \theta) = \sum_{x \in \mathcal{X}(y)} \frac{p(x \mid \theta)}{p(y \mid \theta)} \sum_{v \in \tilde{v}} \chi^{\pi_{\tilde{v}}}(x_{pa(v)}) \left(t_{\tilde{v}|\pi_{\tilde{v}}}(x_v) - \tau(\theta_{\tilde{v}|\pi_{\tilde{v}}})\right).$$

Applying that

$$p(x \mid y, \theta) = \begin{cases} \frac{p(x \mid \theta)}{p(y \mid \theta)} & \text{for } x \in \mathcal{X}(y) \text{ and } p(y \mid \theta) > 0 \\ 0 & \text{otherwise.} \end{cases} \quad (9)$$

we hereby obtain the final expression for local components of the score

$$S_{\tilde{v}|\pi_{\tilde{v}}}(y \mid \theta) = \sum_{v \in \tilde{v}} \sum_{i_{fa(v)} \in \mathcal{I}_{fa(v)}} \left[ p(i_{fa(v)} \mid y, \theta) \right. \\ \left. \times \chi^{\pi_{\tilde{v}}}(i_{pa(v)}) \left(t_{\tilde{v}|\pi_{\tilde{v}}}(i_v) - \tau(\theta_{\tilde{v}|\pi_{\tilde{v}}})\right) \right], \quad (10)$$

where $fa(v)$ is a short notation for the family $v \cup pa(v)$.

The Lauritzen-Spiegelhalter (L-S) procedure for probability propagation (Lauritzen and Spiegelhalter 1988) can be used as an efficient method for calculating the posterior probabilities $p(i_{fa(v)} \mid y, \theta)$. A concise description of this dedication of the L-S procedure can be found in Lauritzen (1995). The remaining parts of (10) are either directly extracted or easily calculated from the exponential representation of the local model.

*Example (continued):* Consider a situation where incompleteness is caused by an imprecise observation. Say, that $X_6$ has four possible values $\mathcal{I}_6 = (i^0, i^1, i^2, i^3)$ for which the collector of data could not decide on one of the two values $x_6 = i^2$ or $x_6 = i^3$. In this case $\mathcal{X}(y) = \{(x_1, x_2, x_3, x_4, x_5, i^2), (x_1, x_2, x_3, x_4, x_5, i^3)\}$. As $p(X_6 \mid x_5, \theta) = (p^0, p^1, p^2, p^3)$, the support of $p(X_5, X_6 \mid y, \theta)$ is given by the non-zero values $(\frac{p^2}{p^2+p^3}, \frac{p^3}{p^2+p^3})$ obtained for $(X_5, X_6) = (x_5, i^2)$ and $(X_5, X_6) = (x_5, i^3)$. Realize also that $\tau(\theta) = (p^1, p^2, p^3)$. By equation (10), the local score is then easily calculated as

$$S_{6|x_5}(y \mid \theta) = \frac{p^2}{p^2+p^3}(-p^1, 1-p^2, -p^3) + \frac{p^3}{p^2+p^3}(-p^1, -p^2, 1-p^3).$$

$\square$

## 4 Observed information in incomplete observation

Let $I(y \mid \theta) = -\frac{\partial^2}{\partial \theta^2} \log p(y \mid \theta)$ denote the observed information in the incomplete observation $y$. We divide the information into local information matrices of dimension $|\Theta_{\tilde{u}|\pi_{\tilde{u}}}| \times |\Theta_{\tilde{v}|\pi_{\tilde{v}}}|$, where each local matrix represents the part of the information matrix as defined by the local components $\theta_{\tilde{u}|\pi_{\tilde{u}}}$ and $\theta_{\tilde{v}|\pi_{\tilde{v}}}, \tilde{u}, \tilde{v} \in \tilde{V}$. Consider the local information

$$\begin{aligned} &I_{\tilde{u}|\pi_{\tilde{u}}, \tilde{v}|\pi_{\tilde{v}}}(y \mid \theta) \\ &= -\frac{\partial^2}{\partial\theta_{\tilde{u}|\pi_{\tilde{u}}}\partial\theta_{\tilde{v}|\pi_{\tilde{v}}}} \log p(y \mid \theta) \\ &= S_{\tilde{u}|\pi_{\tilde{u}}}(y \mid \theta) S_{\tilde{v}|\pi_{\tilde{v}}}(y \mid \theta) \\ &\quad - \frac{1}{p(y \mid \theta)} \sum_{x \in \mathcal{X}(y)} \frac{\partial^2}{\partial\theta_{\tilde{u}|\pi_{\tilde{u}}}\partial\theta_{\tilde{v}|\pi_{\tilde{v}}}} p(x \mid \theta). \end{aligned} \quad (11)$$

The first term in (11) is just the product of local scores. Hence, the aim is now to derive a calculable expression for the second term. Define the Kronecker delta

$$\delta_{\tilde{u}|\pi_{\tilde{u}}, \tilde{v}|\pi_{\tilde{v}}} = \begin{cases} 1 & \text{for } \tilde{u} = \tilde{v} \text{ and } \pi_{\tilde{u}} = \pi_{\tilde{v}} \\ 0 & \text{otherwise}. \end{cases}$$

By straightforward differentiation of (8) with respect to the local component $\theta_{\tilde{u}|\pi_{\tilde{u}}}$

$$\begin{aligned} &\frac{\partial^2}{\partial\theta_{\tilde{u}|\pi_{\tilde{u}}}\partial\theta_{\tilde{v}|\pi_{\tilde{v}}}} p(x \mid \theta) \\ &= p(x \mid \theta) \Bigg( \sum_{u \in \tilde{u}} \chi^{\pi_{\tilde{u}}}(x_{pa(u)}) \big(t_{\tilde{u}|\pi_{\tilde{u}}}(x_u) - \tau(\theta_{\tilde{u}|\pi_{\tilde{u}}})\big) \\ &\quad \times \sum_{v \in \tilde{v}} \chi^{\pi_{\tilde{v}}}(x_{pa(v)}) \big(t_{\tilde{v}|\pi_{\tilde{v}}}(x_v) - \tau(\theta_{\tilde{v}|\pi_{\tilde{v}}})\big) \\ &\quad - \delta_{\tilde{u}|\pi_{\tilde{u}}, \tilde{v}|\pi_{\tilde{v}}} \sum_{v \in \tilde{v}} \chi^{\pi_{\tilde{v}}}(x_{pa(v)}) \nu(\theta_{\tilde{v}|\pi_{\tilde{v}}}) \Bigg), \end{aligned}$$

where $\nu(\theta_{\tilde{v}|\pi_{\tilde{v}}}) = \mathbf{V}[t_{\tilde{v}|\pi_{\tilde{v}}}(X_{\tilde{v}}) \mid \pi_{\tilde{v}}, \theta_{\tilde{v}|\pi_{\tilde{v}}}]$, the covariance matrix for the canonical statistic in the exponential representation of $p(\cdot \mid \pi_{\tilde{v}}, \theta_{\tilde{v}|\pi_{\tilde{v}}})$.

Using (9), the second part of the local information is hereby derived as

$$\begin{aligned} &\frac{1}{p(y \mid \theta)} \sum_{x \in \mathcal{X}(y)} \frac{\partial^2}{\partial\theta_{\tilde{u}|\pi_{\tilde{u}}}\partial\theta_{\tilde{v}|\pi_{\tilde{v}}}} p(x \mid \theta) \\ &= \sum_{u \in \tilde{u}} \sum_{v \in \tilde{v}} \sum_{i_{fa(u) \cup fa(v)}} \Big[ p(i_{fa(u) \cup fa(v)} \mid y, \theta) \\ &\quad \times \chi^{\pi_{\tilde{u}}}(i_{pa(u)}) \chi^{\pi_{\tilde{v}}}(i_{pa(v)}) \\ &\quad \times \big(t_{\tilde{u}|\pi_{\tilde{u}}}(i_u) - \tau(\theta_{\tilde{u}|\pi_{\tilde{u}}})\big) \big(t_{\tilde{v}|\pi_{\tilde{v}}}(i_v) - \tau(\theta_{\tilde{v}|\pi_{\tilde{v}}})\big)' \Big] \\ &\quad - \delta_{\tilde{u}|\pi_{\tilde{u}}, \tilde{v}|\pi_{\tilde{v}}} \sum_{v \in \tilde{v}} \sum_{i_{pa(v)}} p(i_{pa(v)} \mid y, \theta) \chi^{\pi_{\tilde{v}}}(i_{pa(v)}) \nu(\theta_{\tilde{v}|\pi_{\tilde{v}}}). \end{aligned}$$

(12)

By realizing that

$$p(i_{fa(u) \cup fa(v)} \mid y, \theta) = p(i_{fa(u)} \mid i_{fa(v)}, y, \theta) p(i_{fa(v)} \mid y, \theta),$$

the dedication of the L-S procedure can be used for the calculation of the posterior probabilities in (12). The remaining part is directly extracted or easily calculated from the local exponential models. Hence, a final calculable expression for local informations can now be obtained by inserting (10) and (12) into (11).

To discuss the effect that incompleteness of data imposes on structural characteristics for the observed information we reorganize the final expression as

$$\begin{aligned} &I_{\tilde{u}|\pi_{\tilde{u}}, \tilde{v}|\pi_{\tilde{v}}}(y \mid \theta) \\ &= \sum_{u \in \tilde{u}} \sum_{v \in \tilde{v}} \sum_{i_{fa(u) \cup fa(v)}} \Big[ \Big( p(i_{fa(u)} \mid y, \theta) p(i_{fa(v)} \mid y, \theta) \\ &\quad - p(i_{fa(u) \cup fa(v)} \mid y, \theta) \Big) \chi^{\pi_{\tilde{u}}}(i_{pa(u)}) \chi^{\pi_{\tilde{v}}}(i_{pa(v)}) \\ &\quad \times \big(t_{\tilde{u}|\pi_{\tilde{u}}}(i_u) - \tau(\theta_{\tilde{u}|\pi_{\tilde{u}}})\big) \big(t_{\tilde{v}|\pi_{\tilde{v}}}(i_v) - \tau(\theta_{\tilde{v}|\pi_{\tilde{v}}})\big)' \Big] \\ &\quad + \delta_{\tilde{u}|\pi_{\tilde{u}}, \tilde{v}|\pi_{\tilde{v}}} \sum_{v \in \tilde{v}} \sum_{i_{pa(v)}} p(i_{pa(v)} \mid y, \theta) \chi^{\pi_{\tilde{v}}}(i_{pa(v)}) \nu(\theta_{\tilde{v}|\pi_{\tilde{v}}}). \end{aligned}$$

(13)

Consider a complete observation $x$. The posterior probabilities on a subset $X_A, A \subseteq V$ of variables are then given by

$$p(i_A \mid x, \theta) = \begin{cases} 1 & \text{for } i_A = x_A \\ 0 & \text{otherwise.} \end{cases} \quad (14)$$

In this case the local information in (13) reduces to

$$I_{\tilde{u}\mid\pi_{\tilde{u}},\tilde{v}\mid\pi_{\tilde{v}}}(x \mid \theta) = \delta_{\tilde{u}\mid\pi_{\tilde{u}},\tilde{v}\mid\pi_{\tilde{v}}} \sum_{i_{pa(v)}} \nu(\theta_{\tilde{v}\mid\pi_{\tilde{v}}}).$$

Hence, for a complete observation the information matrix will be block-diagonal on local components of the parameter vector. For an incomplete observation this is not the case. The fact that the adjustment of the information due to incompleteness of the observation will undermine the block-diagonality is easily seen by realizing that (14) is no longer valid.

*Example (continued)*: Consider the local information $I_{6\mid x_5, 6\mid x_5}(y \mid \theta)$. Given a complete observation $y = x = (x_1, x_2, x_3, x_4, x_5, x_6)$, the local information equals

$$I_{6\mid x_5, 6\mid x_5}(x \mid \theta) = \nu(\theta_{6\mid x_5})$$
$$= \begin{bmatrix} p^1 - p^1 p^1 & -p^1 p^2 & -p^1 p^3 \\ -p^2 p^1 & p^2 - p^2 p^2 & -p^2 p^3 \\ -p^3 p^1 & -p^3 p^2 & p^3 - p^3 p^3 \end{bmatrix}$$

Actually, this result only depends on the fact that $x_5$ and $x_6$ are observed.

Now, say that $x_6$ was not observed. In this case, the first part of the local information, as given by the product of scores in (11), is 0. Calculations on (12) show that the second part equals $\nu(\theta_{6\mid x_5}) - \nu(\theta_{6\mid x_5}) = 0$. Hence, the observed local information is 0. This is in agreement with the fact that we do not have any information about the conditional distribution $p(X_6 \mid x_5, \theta_{6\mid x_5})$, when $X_6$ is unobserved.

We emphasize that this result is a concequence of the fact that the non-zero values for $p(X_{fa(v)} \mid y, \theta)$ equals $p(X_v \mid x_{pa(v)}, \theta)$, which is not true in general if $X_v$ has observed descendants.  □

## 5   Sample score and observed information

Let $\mathbf{y} = (y^1, y^2, \ldots, y^L)$ denote a sample of possibly incomplete observations which are mutually independent. The likelihood then factorizes over each observation

$$p(\mathbf{y} \mid \theta) \propto \prod_{l=1}^{L} p(y^l \mid \theta) = \prod_{l=1}^{L} \sum_{x^l \in \mathcal{X}(y^l)} p(x^l \mid \theta).$$

As the likelihood is proportional to the product of likelihoods for the individual observations, the sample score and observed sample information are obtained by simply adding the individual scores and informations, respectively. Denote by $n^*(i_A) = \sum_{l=1}^{L} p(i_A \mid y^l, \theta)$ the expected marginal count of observations for the marginal configuration of levels $i_A \in \mathcal{I}_A$. By adding the local scores, as given in (10), the sample score $S(\mathbf{y} \mid \theta)$ has local components

$$S_{\tilde{v}\mid\pi_{\tilde{v}}}(\mathbf{y} \mid \theta) = \sum_{v \in \tilde{v}} \sum_{i_{fa(v)}} \left[ n^*(i_{fa(v)}) \right.$$
$$\left. \times \chi^{\pi_{\tilde{v}}}(i_{pa(v)}) \left(t_{\tilde{v}\mid\pi_{\tilde{v}}}(i_v) - \tau(\theta_{\tilde{v}\mid\pi_{\tilde{v}}})\right) \right] (15)$$

Similarly, by adding the observed informations, the observed sample information $I(\mathbf{y} \mid \theta)$ has local components

$$I_{\tilde{u}\mid\pi_{\tilde{u}},\tilde{v}\mid\pi_{\tilde{v}}}(\mathbf{y} \mid \theta)$$
$$= \sum_{l=1}^{L} \left( \sum_{u \in \tilde{u}} \sum_{i_{fa(u)}} \left[ p(i_{fa(u)} \mid y^l, \theta) \right. \right.$$
$$\left. \times \chi^{\pi_{\tilde{u}}}(i_{pa(u)}) \left(t_{\tilde{u}\mid\pi_{\tilde{u}}}(i_u) - \tau(\theta_{\tilde{u}\mid\pi_{\tilde{u}}})\right) \right]$$
$$\times \sum_{v \in \tilde{v}} \sum_{i_{fa(v)}} \left[ p(i_{fa(v)} \mid y^l, \theta) \right.$$
$$\left. \left. \times \chi^{\pi_{\tilde{v}}}(i_{pa(v)}) \left(t_{\tilde{v}\mid\pi_{\tilde{v}}}(i_v) - \tau(\theta_{\tilde{v}\mid\pi_{\tilde{v}}})\right)' \right] \right)$$
$$- \sum_{u \in \tilde{u}} \sum_{v \in \tilde{v}} \sum_{i_{fa(v) \cup fa(u)}} \left[ n^*(i_{fa(v) \cup fa(u)}) \right.$$
$$\times \chi^{\pi_{\tilde{u}}}(i_{pa(u)}) \chi^{\pi_{\tilde{v}}}(i_{pa(v)})$$
$$\left. \times \left(t_{\tilde{u}\mid\pi_{\tilde{u}}}(i_u) - \tau(\theta_{\tilde{u}\mid\pi_{\tilde{u}}})\right) \left(t_{\tilde{v}\mid\pi_{\tilde{v}}}(i_v) - \tau(\theta_{\tilde{v}\mid\pi_{\tilde{v}}})\right)' \right]$$
$$+ \delta_{\tilde{u}\mid\pi_{\tilde{u}},\tilde{v}\mid\pi_{\tilde{v}}} \sum_{v \in \tilde{v}} \sum_{i_{pa(v)}} n^*(i_{pa(v)}) \chi^{\pi_{\tilde{v}}}(i_{pa(v)}) \nu(\theta_{\tilde{v}\mid\pi_{\tilde{v}}}).$$
$$(16)$$

The expression (16) is organized so that the first term should be easy to identify as the sum of the products of local scores for each observation. Hence, in case the local scores of each observation have already been calculated, one might replace the first term of (16) by

$$\sum_{l=1}^{L} \left(S_{\tilde{u}\mid\pi_{\tilde{u}}}(y^l \mid \theta) S_{\tilde{v}\mid\pi_{\tilde{v}}}(y^l \mid \theta)'\right).$$

Notice that a lot of posterior probabilities $p(i_A \mid y, \theta)$, $A \in V$ have to be calculated to complete the calculations. Hence computational efficiency demands an efficient method for calculating these, as e.g. the L-S procedure.

## 6  Posterior score and observed information

Suppose that we have information about $\theta$ in the form of a prior distribution of parameters considered as random variables, $p(\theta)$. The posterior distribution given an incomplete sample (or single observation) is then defined as

$$p(\theta \,|\, \mathbf{y}) = \frac{p(\mathbf{y} \,|\, \theta) p(\theta)}{p(\mathbf{y})}. \tag{17}$$

Here we consider a Bayesian interpretation of the score and information. In analogy with the traditional score and observed information, let $S(\theta \,|\, \mathbf{y}) = \frac{\partial}{\partial \theta} \log p(\theta \,|\, \mathbf{y})$ and $I(\theta \,|\, \mathbf{y}) = -\frac{\partial^2}{\partial \theta^2} \log p(\theta \,|\, \mathbf{y})$ be denoted as the posterior score and observed information, and let $S(\theta) = \frac{\partial}{\partial \theta} \log p(\theta)$ and $I(\theta) = -\frac{\partial^2}{\partial \theta^2} \log p(\theta)$ be denoted as the prior score and information. From (17) it is easily seen that the posterior score and information are obtained by simply adding, respectively, the traditional score and information onto the prior score and information. Hence,

$$S(\theta \,|\, \mathbf{y}) = S(\mathbf{y} \,|\, \theta) + S(\theta) \tag{18}$$

and

$$I(\theta \,|\, \mathbf{y}) = I(\mathbf{y} \,|\, \theta) + I(\theta). \tag{19}$$

Now, consider the prior distribution of parameters. The construction simplifies considerably by matching assumptions of variation independence with independence of the parameters considered as random variables. Hence, by relaxed *global independence* we assume that $\theta_{\tilde{v}}, \tilde{v} \in \bar{V}$ are mutually independent, and by *local independence* we assume that local components $\theta_{\tilde{v}|\pi_{\tilde{v}}}, \pi_{\tilde{v}} \in \mathcal{I}_{pa(\tilde{v})}$ are mutually independent. The notion of global and local independence can also be found in Spiegelhalter and Lauritzen (1990a, 1990b) and the line of work as reported in Heckerman et al. (1995).

Under the assumptions of relaxed global and local independence the distribution of parameters factorizes as

$$p(\theta) = \prod_{\tilde{v} \in \bar{V}} \prod_{\pi_{\tilde{v}} \in \mathcal{I}_{pa(\tilde{v})}} p(\theta_{\tilde{v}|\pi_{\tilde{v}}}).$$

By this factorization the local components for the prior score

$$S_{\tilde{v}|\pi_{\tilde{v}}}(\theta) = \frac{\partial}{\partial \theta_{\tilde{v}|\pi_{\tilde{v}}}} \log p(\theta_{\tilde{v}|\pi_{\tilde{v}}})$$

and the local components for the prior information

$$I_{\tilde{u}|\pi_{\tilde{u}}, \tilde{v}|\pi_{\tilde{v}}}(\theta) = -\delta_{\tilde{u}|\pi_{\tilde{u}}, \tilde{v}|\pi_{\tilde{v}}} \frac{\partial^2}{\partial \theta_{\tilde{v}|\pi_{\tilde{v}}}^2} \log p(\theta_{\tilde{v}|\pi_{\tilde{v}}})$$

are derived from local prior distributions.

## 7  Specific prior distributions on parameters

To complete the specification for the posterior score and observed information, we consider parametrizations for the prior distribution of parameters.

Intending a unification of a batch method for quantifying probabilistic expert systems by the mode which maximizes the posterior distribution, as described in Thiesson (1995), and a method for sequential updating of conditional probabilities, as described in Spiegelhalter and Lauritzen (1990a, 1990b), we are especially interested in conjugate distributions (or approximately conjugate). By the intended unification, a system can be initialized by the batch learning method, and following, as new data accumulates, the system can be updated and improved by the sequential updating method.

### 7.1  Conjugate prior for local exponential models

For the general setting we consider the prior distribution of parameters as a member of the conjugate model for the likelihood as defined by a recursive exponential model.

Let $\mathbf{x} = (x^1, \ldots, x^n)$ denote a sample of $n$ complete observations. The observed count for configuration $i_A \in \mathcal{I}_A, A \subseteq V$ is then defined as

$$n(i_A) = \sum_{l=1}^{n} \chi^{i_A}(x^l),$$

where

$$\chi^{i_A}(x^l) = \begin{cases} 1 & \text{for } x^l_A = i_A \\ 0 & \text{otherwise.} \end{cases}$$

For independent observations and by the assumptions of variation independence the likelihood factorizes as

$$p(\mathbf{x} \,|\, \theta) \propto \prod_{i \in \mathcal{I}} p(i \,|\, \theta)^{n(i)}$$
$$= \prod_{\tilde{v} \in \bar{V}} \prod_{v \in \tilde{v}} \prod_{\pi_v \in \mathcal{I}_{pa(v)}} \prod_{i_v \in \mathcal{I}_v} p(i_v \,|\, \pi_v, \theta_{\tilde{v}|\pi_{\tilde{v}}})^{n(i_v, \pi_v)}$$
$$= \prod_{\tilde{v} \in \bar{V}} \prod_{\pi_{\tilde{v}} \in \mathcal{I}_{pa(\tilde{v})}} \prod_{i_{\tilde{v}} \in \mathcal{I}_{\tilde{v}}} p(i_{\tilde{v}} \,|\, \pi_{\tilde{v}}, \theta_{\tilde{v}|\pi_{\tilde{v}}})^{\sum_{v \in \tilde{v}} n(i_v, \pi_v)},$$

where the last equality follows from the fact that conditional probability tables are equal for all $v \in \tilde{v}$. We observe that the likelihood factorizes into a product of local likelihoods

$$p_{\tilde{v}|\pi_{\tilde{v}}}(\mathbf{x} \,|\, \theta_{\tilde{v}|\pi_{\tilde{v}}}) = \prod_{i_{\tilde{v}} \in \mathcal{I}_{\tilde{v}}} p(i_{\tilde{v}} \,|\, \pi_{\tilde{v}}, \theta_{\tilde{v}|\pi_{\tilde{v}}})^{\sum_{v \in \tilde{v}} n(i_v, \pi_v)},$$

460  Thiesson

each defined by the exponential model (4). The natural local conjugate priors are therefore defined by conjugate exponential models (Diaconis and Ylvisaker 1979)

$$p(\theta_{\tilde{v}|\pi_{\tilde{v}}}) \propto \exp\left(\theta'_{\tilde{v}|\pi_{\tilde{v}}}\kappa - \beta\phi(\theta_{\tilde{v}|\pi_{\tilde{v}}})\right),$$

where $\kappa$ is a vector of same dimension as $\theta_{\tilde{v}|\pi_{\tilde{v}}}$ and $\beta$ is a scalar.

Let $\theta^*_{\tilde{v}|\pi_{\tilde{v}}}$ denote the value which maximizes $p(\theta_{\tilde{v}|\pi_{\tilde{v}}})$. In the neighbourhood of $\theta^*_{\tilde{v}|\pi_{\tilde{v}}}$ a Taylor series expansion implies that

$$\log p(\theta_{\tilde{v}|\pi_{\tilde{v}}}) \approx$$
$$\log p(\theta^*_{\tilde{v}|\pi_{\tilde{v}}}) - \frac{\beta}{2}(\theta_{\tilde{v}|\pi_{\tilde{v}}} - \theta^*_{\tilde{v}|\pi_{\tilde{v}}})'\nu(\theta^*_{\tilde{v}|\pi_{\tilde{v}}})(\theta_{\tilde{v}|\pi_{\tilde{v}}} - \theta^*_{\tilde{v}|\pi_{\tilde{v}}}).$$

Hence, a conjugate local prior is approximately proportional to the multivariate normal distribution $\mathcal{N}\left(\theta^*_{\tilde{v}|\pi_{\tilde{v}}}, \frac{1}{\beta}\nu(\theta^*_{\tilde{v}|\pi_{\tilde{v}}})^{-1}\right)$ in a neighbourhood of $\theta^*_{\tilde{v}|\pi_{\tilde{v}}}$.

Local prior scores for this approximation are derived as

$$S_{\tilde{v}|\pi_{\tilde{v}}}(\theta) = -\beta\nu(\theta^*_{\tilde{v}|\pi_{\tilde{v}}})(\theta_{\tilde{v}|\pi_{\tilde{v}}} - \theta^*_{\tilde{v}|\pi_{\tilde{v}}}),$$

and local prior informations as

$$I_{\tilde{u}|\pi_{\tilde{u}},\tilde{v}|\pi_{\tilde{v}}}(\theta) = \delta_{\tilde{u}|\pi_{\tilde{u}},\tilde{v}|\pi_{\tilde{v}}}\beta\nu(\theta^*_{\tilde{v}|\pi_{\tilde{v}}}).$$

The mean $\theta^*_{\tilde{v}|\pi_{\tilde{v}}}$ and the parameter $\beta$ are unknown factors of the approximately conjugate multivariate normal distribution, which have to be extracted from expert knowledge. For this task it seems reasonable to request domain experts to give a "best guess" $\hat{p}(X_v \mid i_{pa(v)})$ on each conditional distribution with an assessment of imprecision (or confidence) on each of the probabilities in the form of an interval of variation. Here it should be noticed that if the expert specification is not the same for each table of conditional distributions, $\hat{p}(X_v \mid X_{pa(v)})$, where $v \in \tilde{v}$, these should be forced equal or the expert should reconsider the equality of these tables. If the partitioning into equal tables is indisputable one should only request for generic "best guess" tables $\hat{p}(X_{\tilde{v}} \mid X_{pa(\tilde{v})}), \tilde{v} \in \tilde{V}$ with assessments of imprecision.

The mean $\theta^*_{\tilde{v}|\pi_{\tilde{v}}}$ is derived as the value of $\theta_{\tilde{v}|\pi_{\tilde{v}}}$ which minimizes the Kullback-Leibler discrepancy between $\hat{p}(X_{\tilde{v}} \mid \pi_{\tilde{v}})$ and $p(X_{\tilde{v}} \mid \pi_{\tilde{v}}, \theta_{\tilde{v}|\pi_{\tilde{v}}})$

$$KL\left(\hat{p}(X_{\tilde{v}} \mid \pi_{\tilde{v}}), p(X_{\tilde{v}} \mid \pi_{\tilde{v}}, \theta_{\tilde{v}|\pi_{\tilde{v}}})\right)$$
$$= \sum_{i_{\tilde{v}} \in \mathcal{I}_{\tilde{v}}} \hat{p}(i_{\tilde{v}} \mid \pi_{\tilde{v}}) \log \frac{\hat{p}(i_{\tilde{v}} \mid \pi_{\tilde{v}})}{p(i_{\tilde{v}} \mid \pi_{\tilde{v}}, \theta_{\tilde{v}|\pi_{\tilde{v}}})}.$$

Here, we use the convention $0\log(0/a) = 0$ for $a \geq 0$ and $a\log(a/0) = \infty$ for $a > 0$.

The first and second order derivatives of the discrepancy can be found as respectively

$$\frac{\partial}{\partial \theta_{\tilde{v}|\pi_{\tilde{v}}}} KL\left(\hat{p}(X_{\tilde{v}} \mid \pi_{\tilde{v}}), p(X_{\tilde{v}} \mid \pi_{\tilde{v}}, \theta_{\tilde{v}|\pi_{\tilde{v}}})\right)$$
$$= -\sum_{i_{\tilde{v}} \in \mathcal{I}_{\tilde{v}}} \hat{p}(i_{\tilde{v}} \mid \pi_{\tilde{v}})\left(t_{\tilde{v}|\pi_{\tilde{v}}}(i_{\tilde{v}}) - \tau(\theta_{\tilde{v}|\pi_{\tilde{v}}})\right)$$

and

$$\frac{\partial^2}{\partial \theta_{\tilde{u}|\pi_{\tilde{u}}}\partial \theta_{\tilde{v}|\pi_{\tilde{v}}}} KL\left(\hat{p}(X_{\tilde{v}} \mid \pi_{\tilde{v}}), p(X_{\tilde{v}} \mid \pi_{\tilde{v}}, \theta_{\tilde{v}|\pi_{\tilde{v}}})\right)$$
$$= \delta_{\tilde{u}|\pi_{\tilde{u}},\tilde{v}|\pi_{\tilde{v}}}\nu(\theta_{\tilde{v}|\pi_{\tilde{v}}}).$$

Besides the "best guess" these expressions do not involve statistics which are not already in demand for the implementation of the traditional score and observed information. Hence, with little additional effort in implementation, the minimizing $\theta^*_{\tilde{v}|\pi_{\tilde{v}}}$ can be found numerically (if not analytically) by e.g. a Newton-Raphson method.

Ideally the discrepancy is 0. In situations, however, the discrepancy may be non-zero, which reflects inconsistency between the specified "best guess" distribution and the structural restrictions as specified for the distribution. In this case we choose the nearest distribution (by the discrepancy), which obeys the restrictions. If the discrepancy is very large, this should affect the confidence in the specified distribution or the restrictions.

The parameter $\beta$ that adjusts the variance for the multivariate normal distribution is determined from the intervals of variation as specified for each probability. By assuming that an interval of variation for a probability equals twice the standard derivation for the marginal distribution of the probability, the adjustment factor is derived as follows.

Let $SD(i_{\tilde{v}} \mid \pi_{\tilde{v}})$ denote half the interval of variation for $p(i_{\tilde{v}} \mid \pi_{\tilde{v}}, \theta_{\tilde{v}|\pi_{\tilde{v}}})$, and denote by $\mathbf{V}\left(p(i_{\tilde{v}} \mid \pi_{\tilde{v}}, \theta_{\tilde{v}|\pi_{\tilde{v}}})\right)$ the variance of that probability. By the delta method the variance matrix for probabilities can be approximated from the variance matrix for parameters. If we consider the variances for the marginal distributions of probabilities only, these are approximated by the diagonal elements

$$\mathbf{V}\left(p(i_{\tilde{v}} \mid \pi_{\tilde{v}}, \theta_{\tilde{v}|\pi_{\tilde{v}}})\right)$$
$$= \frac{\partial p(i_{\tilde{v}} \mid \pi_{\tilde{v}}, \theta_{\tilde{v}|\pi_{\tilde{v}}})}{\partial \theta^*_{\tilde{v}|\pi_{\tilde{v}}}}' \frac{1}{\beta}\nu(\theta^*_{\tilde{v}|\pi_{\tilde{v}}})^{-1} \frac{\partial p(i_{\tilde{v}} \mid \pi_{\tilde{v}}, \theta_{\tilde{v}|\pi_{\tilde{v}}})}{\partial \theta^*_{\tilde{v}|\pi_{\tilde{v}}}},$$
(20)

where

$$\frac{\partial p(i_{\tilde{v}} \mid \pi_{\tilde{v}}, \theta_{\tilde{v}|\pi_{\tilde{v}}})}{\partial \theta^*_{\tilde{v}|\pi_{\tilde{v}}}} = p(i_{\tilde{v}} \mid \pi_{\tilde{v}}, \theta^*_{\tilde{v}|\pi_{\tilde{v}}})\left(t_{\tilde{v}|\pi_{\tilde{v}}}(i_{\tilde{v}}) - \tau(\theta^*_{\tilde{v}|\pi_{\tilde{v}}})\right).$$

The adjustment factor $\beta(i_{\tilde{v}})$, associated for the marginal distribution of $p(i_{\tilde{v}} \mid \pi_{\tilde{v}}, \theta_{\tilde{v}\mid\pi_{\tilde{v}}})$, can now be derived from equation (20) by utilizing that $SD(i_{\tilde{v}} \mid \pi_{\tilde{v}})^2 = \mathbf{V}\left(p(i_{\tilde{v}} \mid \pi_{\tilde{v}}, \theta_{\tilde{v}\mid\pi_{\tilde{v}}})\right)$. In case of inconsistency between the calculated adjustment factors for different $i_{\tilde{v}} \in \mathcal{I}_{\tilde{v}}$ we choose the factor which implies the lowest precision for the normal distribution. Hence,

$$\beta = \min\{\beta(i_{\tilde{v}}) \mid i_{\tilde{v}} \in \mathcal{I}_{\tilde{v}}\}.$$

*Example (continued):* Assume that the local distribution $p(X_6 \mid \pi_6, \xi, \gamma) = (p^0, p^1, p^2, p^3)$ is restricted by the log-linear form $\log p(X_6 \mid \pi_6, \xi, \gamma) = \xi + X_6 \gamma$, where the levels for $X_6$ are real-valued quantities, say $\mathcal{I}_6 = (i^0, i^1, i^2, i^3) = (0, 1, 2, 3)$.

Let $\eta_{6\mid\pi_6}$ and $s_{6\mid\pi_6}(X_6)$ denote the parameter vector and the statistic for the exponential representation of the local distribution without structural restrictions, as derived earlier in this example (page 3). Then $p(X_6 \mid \pi_6, \xi, \gamma)$ can be represented as the distribution $p(X_6 \mid \pi_6, \theta_{6\mid\pi_6})$ formed by the exponential sub-model of order 1, with parameter $\theta_{6\mid\pi_6} = \gamma$ given by the affine transformation $\eta_{6\mid\pi_6} = (\log\frac{p^1}{p^0}, \log\frac{p^2}{p^0}, \log\frac{p^3}{p^0}) = T\theta_{6\mid\pi_6}$, where $T' = (i^1 - i^0, i^2 - i^0, i^3 - i^0) = (1, 2, 3)$. For this representation $t_{6\mid\pi_6}(X_6) = T's_{6\mid\pi_6}(X_6)$, $\tau(\theta_{6\mid\pi_6}) = T'\tau(\eta_{6\mid\pi_6})$, and $\nu(\theta_{6\mid\pi_6}) = T'\nu(\eta_{6\mid\pi_6})T$.

Now, say that the opinion about the local distribution has been imprecisely specified as $(0.05[0.02 - 0.08], 0.10[0.05 - 0.15], 0.25[0.20 - 0.30], 0.60[0.50 - 0.70])$, where each interval denotes the imprecision of the "best guess" in front of it.

The "best guess", $\hat{p}(X_v \mid x_{pa(v)}) = (0.05, 0.10, 0.25, 0.60)$, almost satisfies the structural restriction. However, a simple check reveals that log-odds disagree on the value for $\theta_{6\mid\pi_6}$ which parameterizes the exponential representation of the distribution. Therefore we use a Newton-Raphson method to determine the parameter value $\theta^*_{6\mid\pi_6}$, which implies the lowest KL-discrepancy between the "best guess" and a distribution of the correct functional form. For this task, first and second order derivatives are derived as $\frac{\partial}{\partial \theta_{6\mid\pi_6}}KL = (p^1 + 2p^2 + 3p^3) - 2.4$ and $\frac{\partial^2}{\partial \theta^2_{6\mid\pi_6}}KL = (p^1 + 4p^2 + 9p^3) - (p^1 + 2p^2 + 3p^3)(p^1 + 2p^2 + 3p^3)$.

Now, starting from an initial value $\theta_{6\mid\pi_6} = \log\frac{p^1}{p^0} = \log 2$, five Newton-Raphson iterations given by

$$\theta_{6\mid\pi_6} \leftarrow \theta_{6\mid\pi_6} - \left(\frac{\partial^2}{\partial \theta^2_{6\mid\pi_6}}KL\right)^{-1}\frac{\partial}{\partial \theta_{6\mid\pi_6}}KL$$

move the parameter value into an acceptable value $\theta^*_{6\mid\pi_6} = 0.86148$, as displayed in table 1.

Consider now the variance for the approximate conjugate normal distribution. Except from the adjustment $1/\beta$, it falls right out of the last Newton-Raphson iteration as $\left(\frac{\partial^2}{\partial(\theta^*_{6\mid\pi_6})^2}KL\right)^{-1} = 1.3832$.

Let $p^{*r}$ denote the probability $p(i^r \mid \pi_6, \theta^*_{6\mid\pi_6})$, $r = 0, 1, 2, 3$. Then $\frac{\partial p(i^r \mid \pi_6, \theta_{6\mid\pi_6})}{\partial \theta^*_{6\mid\pi_6}} = i^r p^{*r} - p^{*r}(p^{*1} + 2p^{*2} + 3p^{*3})$, which for $r = 0, 1, 2, 3$ equals respectively $-0.10800$, $-0.14908$, $-0.10081$, and $0.35790$. For each interval of variation we can now use (20) to calculate the individual adjustment factors $\beta(i^r), r = 0, 1, 2, 3$, as respectively 17.9, 12.3, 5.6, and 17.7. Being the smallest, 5.6 is chosen as the adjustment factor for the variance.

Hence, the local conjugate distribution $p(\theta_{6\mid\pi_6})$ is approximated by $\mathcal{N}(0.861, 0.247)$.

$\square$

## 7.2 Conjugate prior for local multinomial models

In most real situations some of the local models will not hold structural restrictions. In case of multinomial sampling with respect to variables for a local likelihood the natural local conjugate prior distribution of parameters is defined by a Dirichlet distribution. Consider a generic local distribution and assume that the probabilities a priori are Dirichlet distributed $\mathcal{D}(\alpha(i_{\tilde{v}}, \pi_{\tilde{v}}); i_{\tilde{v}} \in \mathcal{I}_{\tilde{v}})$ with a total number of $|\mathcal{I}_{\tilde{v}}|$ parameters; a parameter for each index of $\mathcal{I}_{\tilde{v}}$. Hence, the prior distribution of probabilities is in the form

$$p\left(p(i_{\tilde{v}} \mid \pi_{\tilde{v}}, \theta_{\tilde{v}\mid\pi_{\tilde{v}}}), i_{\tilde{v}} \in \mathcal{I}_{\tilde{v}}\right) \propto \prod_{i_{\tilde{v}} \in \mathcal{I}_{\tilde{v}}} p(i_{\tilde{v}} \mid \pi_{\tilde{v}}, \theta_{\tilde{v}\mid\pi_{\tilde{v}}})^{\alpha(i_{\tilde{v}}, \pi_{\tilde{v}})-1}.$$

By a transformation, as given by the exponential representation of probabilities, the prior becomes a distribution of $\theta_{\tilde{v}\mid\pi_{\tilde{v}}}$. The distribution is given by (noting that the determinant of the Jacobian for the local distribution $\frac{d}{d\theta_{\tilde{v}\mid\pi_{\tilde{v}}}}p(\cdot \mid \pi_{\tilde{v}}, \theta_{\tilde{v}\mid\pi_{\tilde{v}}})$ is proportional to $\prod_{i_{\tilde{v}} \in \mathcal{I}_{\tilde{v}}} p(i_{\tilde{v}} \mid \pi_{\tilde{v}}, \theta_{\tilde{v}\mid\pi_{\tilde{v}}})$)

$$p(\theta_{\tilde{v}\mid\pi_{\tilde{v}}}) \propto \prod_{i_{\tilde{v}} \in \mathcal{I}_{\tilde{v}}} \exp\left[\left(\theta'_{\tilde{v}\mid\pi_{\tilde{v}}} t_{\tilde{v}\mid\pi_{\tilde{v}}}(i_{\tilde{v}}) - \phi(\theta_{\tilde{v}\mid\pi_{\tilde{v}}})\right)\alpha(i_{\tilde{v}}, \pi_{\tilde{v}})\right] \quad (21)$$

From (21) the local prior score and information are easily derived as

$$S_{\tilde{v}\mid\pi_{\tilde{v}}}(\theta_{\tilde{v}\mid\pi_{\tilde{v}}}) = \sum_{i_{\tilde{v}} \in \mathcal{I}_{\tilde{v}}} \alpha(i_{\tilde{v}}, \pi_{\tilde{v}})\left(t_{\tilde{v}\mid\pi_{\tilde{v}}}(i_{\tilde{v}}) - \tau(\theta_{\tilde{v}\mid\pi_{\tilde{v}}})\right)$$

$$\quad (22)$$

and

$$I_{\tilde{u}\mid\pi_{\tilde{u}}, \tilde{v}\mid\pi_{\tilde{v}}}(\theta_{\tilde{v}\mid\pi_{\tilde{v}}}) = \delta_{\tilde{u}\mid\pi_{\tilde{u}}, \tilde{v}\mid\pi_{\tilde{v}}}\alpha(\pi_{\tilde{v}})\nu(\theta_{\tilde{v}\mid\pi_{\tilde{v}}}), \quad (23)$$

| $\theta$ | $\frac{\partial}{\partial \theta}KL$ | $\frac{\partial^2}{\partial \theta^2}KL$ | $p^0$ | $p^1$ | $p^2$ | $p^3$ |
|---|---|---|---|---|---|---|
| 0.69315 | -0.13334 | 0.86228 | 0.06666 | 0.13333 | 0.26666 | 0.53333 |
| 0.84781 | -0.00998 | 0.73398 | 0.04650 | 0.10854 | 0.25340 | 0.59156 |
| 0.86141 | -0.00007 | 0.72305 | 0.04500 | 0.10650 | 0.25204 | 0.59645 |
| 0.86151 | 0.00002 | 0.72292 | 0.04499 | 0.10649 | 0.25203 | 0.59649 |
| 0.86148 | 0.00000 | 0.72298 | 0.04500 | 0.10649 | 0.25203 | 0.59648 |

Table 1: Newton-Raphson iterates for the example. The first column displays the parameter values for each iteration. The second and third column display the first and second derivatives of the KL discrepancy evaluated at the parameter values. Column four through seven show the associated distributions.

where $\alpha(\pi_{\tilde{v}}) = \sum_{i_{\tilde{v}} \in \mathcal{I}_{\tilde{v}}} \alpha(i_{\tilde{v}}, \pi_{\tilde{v}})$.

The similarity of (22) and (23) with the expressions for the traditional local score and information, (15) and (16), leads to the observation that the prior Dirichlet distribution has the effect of adding its parameters as imaginary counts to get the posterior expressions.

Depending on the domain expert, naturally, it may in situations seem unreasonable to request a prior opinion about local distributions directly in the form of imaginary counts giving the parameters of a Dirichlet distribution. Again we overcome the problem by letting the domain expert specify a "best guess" on local distributions with an interval of variation on each of the probabilities.

As also suggested in Spiegelhalter et al. (1993) and Heckerman et al. (1995), the parameters of a Dirichlet distribution can then be calculated from the expressions of individual means and variances for each random variable of the distribution

$$\mathbf{E}\left[p(i_{\tilde{v}} \mid \pi_{\tilde{v}}, \theta_{\tilde{v}\mid\pi_{\tilde{v}}})\right] = \frac{\alpha(i_{\tilde{v}}, \pi_{\tilde{v}})}{\alpha(\pi_{\tilde{v}})}$$

$$\mathbf{V}\left[p(i_{\tilde{v}} \mid \pi_{\tilde{v}}, \theta_{\tilde{v}\mid\pi_{\tilde{v}}})\right] = \frac{(\alpha(\pi_{\tilde{v}}) - \alpha(i_{\tilde{v}}, \pi_{\tilde{v}}))\,\alpha(i_{\tilde{v}}, \pi_{\tilde{v}})}{\alpha(\pi_{\tilde{v}})^2\,(\alpha(\pi_{\tilde{v}}) + 1)}.$$

Assume for the conditional probability $p(i_{\tilde{v}} \mid \pi_{\tilde{v}}, \theta_{\tilde{v}\mid\pi_{\tilde{v}}})$ that the mean equals the "best guess" $\hat{p}(i_{\tilde{v}} \mid \pi_{\tilde{v}})$ and that the standard deviation is equal to half the interval of variation $SD(i_{\tilde{v}} \mid \pi_{\tilde{v}})$. The equivalent sample size $\alpha^{i_{\tilde{v}}}(\pi_{\tilde{v}})$, associated for the probability $p(i_{\tilde{v}} \mid \pi_{\tilde{v}}, \theta_{\tilde{v}\mid\pi_{\tilde{v}}})$, is now derived as

$$\alpha^{i_{\tilde{v}}}(\pi_{\tilde{v}}) = \frac{(1 - \hat{p}(i_{\tilde{v}} \mid \pi_{\tilde{v}}))\,\hat{p}(i_{\tilde{v}} \mid \pi_{\tilde{v}})}{SD(i_{\tilde{v}} \mid \pi_{\tilde{v}})^2} - 1.$$

In case of very large intervals $\alpha^{i_{\tilde{v}}}(\pi_{\tilde{v}})$ may become negative. This is regarded as a token of non-informative prior knowledge, in which case $\alpha^{i_{\tilde{v}}}(\pi_{\tilde{v}})$ is set to the non-informative sample size 0.

A consistent specification of the prior distribution requires that $\alpha^{i_{\tilde{v}}}(\pi_{\tilde{v}})$ has the same value for all $i_{\tilde{v}} \in \mathcal{I}_{\tilde{v}}$.

In case of inconsistency between the calculated sample sizes from different interval specifications we choose the smallest. Hence,

$$\alpha(\pi_{\tilde{v}}) = \min\{\alpha^{i_{\tilde{v}}}(\pi_{\tilde{v}}) \mid i_{\tilde{v}} \in \mathcal{I}_{\tilde{v}}\}.$$

By the assumption that the means are given by the "best guess" distribution, the parameters of the Dirichlet distribution are then calculated as

$$\alpha(i_{\tilde{v}}, \pi_{\tilde{v}}) = \hat{p}(i_{\tilde{v}} \mid \pi_{\tilde{v}})\alpha(\pi_{\tilde{v}}).$$

*Example (continued):* Say that the opinion about the conditional distribution without structural restrictions $p(X_3 \mid \pi_3, \theta_{\tilde{3}\mid\pi_3}) = p(X_4 \mid \pi_4, \theta_{\tilde{3}\mid\pi_4})$, $\pi_3 = \pi_4$ has been imprecisely specified as $(0.10[0.04 - 0.16], 0.20[0.10 - 0.30], 0.50[0.40 - 0.60], 0.20[0.10 - 0.30])$. For each imprecisely stated conditional probability we calculate the associated equivalent sample size as respectively 24, 15, 24, and 15. Being the smallest, 15 is chosen as the equivalent sample size for the Dirichlet distribution of $p(X_3 \mid \pi_3, \theta_{3\mid\pi_3})$ and $p(X_4 \mid \pi_4, \theta_{4\mid\pi_4})$. The imaginary counts (the parameters that specify the Dirichlet distribution) then become $(1.5, 3.0, 7.5, 3.0)$. □

## 8  Further issues on modelling

An obvious possibility of even more sophisticated modelling is to relax or annul the assumption of local variation independence. Spiegelhalter and Lauritzen (1990a) suggest (for recursive graphical models) an interesting possibility as follows.

Consider the parameter vector for a generic table of local distributions, $\theta_{\tilde{v}}$. This parameter vector is restricted by assuming that each local vector, $\theta_{\tilde{v}\mid\pi_{\tilde{v}}}$, is defined by linear combinations of functions $u_{\tilde{v}}^1(\pi_{\tilde{v}}), \ldots, u_{\tilde{v}}^r(\pi_{\tilde{v}})$ on parent configurations. Hence, for any parent configuration

$$\theta_{\tilde{v}\mid\pi_{\tilde{v}}} = \alpha_{\tilde{v}} u_{\tilde{v}}(\pi_{\tilde{v}}),$$

where $u_{\bar{v}}(\pi_{\bar{v}}) = \left(u_{\bar{v}}^1(\pi_{\bar{v}}), \ldots, u_{\bar{v}}^r(\pi_{\bar{v}})\right)'$ is a vector holding given values of the functions and $\alpha_{\bar{v}}$ is a $|\theta_{\bar{v}|\pi_{\bar{v}}}| \times r$ dimensional matrix of parameters.

Another interesting prospect on modelling would be to extend the recursive exponential model into models, which we denote as block recursive exponential models. These models may evolve from recursive exponential models by allowing that a block (or group) of variables are sitting at each node of the graphical DAG representation of the model. Relations between variables in different blocks are then causal in the direction of arrows, whereas relations between variables within a block are symmetric. A block may contain variables which are not direct causes to any of the variables within a response block, and vice versa.

The class of recursive graphical models is an important specialization of recursive exponential models. Similarly, the block recursive exponential models should admit a specialization into the block recursive graphical models or chain graph models of Lauritzen and Wermuth (1984, 1989), which demands an annulment of local variation independence as proposed above.

## Acknowledgements

The author is grateful to Steffen L. Lauritzen for helpful comments and for inspiring this work by providing a first draft on the derivation of the score for recursive graphical models. Grant support was provided by ESPRIT Basic Research Action, project no. 6156 (DRUMS II) as well as the PIFT programme of the Danish Research Councils.